\documentclass[twocolumn,aps,notitlepage,showpacs, superscriptaddress]{revtex4}

\usepackage{amsfonts}
\usepackage{amsmath}
\usepackage{amssymb}

\newcommand{\trace}{\text{Tr}}
\begin{document}

\author{Line Hjortsh\o j Pedersen}
\email{lhp@phys.au.dk}
\affiliation{Lundbeck Foundation Theoretical
Center for Quantum System Research, Department of Physics and
Astronomy, University of Aarhus,Ny Munkegade, Bld. 1520, DK-8000 \AA
rhus C, Denmark}
\author{Niels Martin M\o ller}
\affiliation{Department of Mathematical Sciences, University of
Aarhus, Ny Munkegade, Bld. 1530, DK-8000 \AA rhus C, Denmark}
\author{Klaus M\o lmer}
\affiliation{Lundbeck Foundation Theoretical Center for Quantum
System Research, Department of Physics and Astronomy, University of
Aarhus,Ny Munkegade, Bld. 1520, DK-8000 \AA rhus C, Denmark}
\title{Fidelity of quantum operations}


\begin{abstract}
We present a derivation and numerous applications of a compact explicit formula for the average
fidelity of a quantum operation on a finite dimensional quantum system. The formula can be applied to
averages over particularly relevant subspaces; it is easily generalized to multi-component systems,
and as a special result, we show that when the same completely positive trace-preserving map is
applied to a large number of qubits with one-bit fidelity $F$ close to unity, the average fidelity of
the operation on the full $K$-bit register scales as $F^{3K/2}$.
\end{abstract}

\maketitle

In quantum information theory one requires a measure for the fidelity of quantum operations such as
quantum computing gates, quantum state storage and retrieval, transmission and teleportation. The
fidelity characterizes the agreement between the actual outcome of the operation and the desired
state, and it will in general depend on the initial state on which the operation is applied. Since the
operation may generally be applied to any arbitrary state, it is natural to average the fidelity over
all pure initial states, chosen uniformly in the system Hilbert space. Derivations have been given in
the literature for the average fidelity for qubit \cite{HHH} and qudit operations \cite{Nielsen}, as
well as for their evaluation as a sum over a properly chosen discrete set of states \cite{BOSBJ},
\cite{BBM}. The latter approach is particularly relevant in connection with Quantum Process Tomography
\cite{NC}, which provides a procedure to deduce the quantum operation acting on a system from
experimental observations. In this Letter we will present and analyze a simple expression for this
average fidelity for the case where the quantum operation is known, e.g., from a solution of the
Schr\"odinger or Master Equation of the system, and we focus on the calculation of the fidelity with
which the known actual operation resembles the desired operation on the quantum system. We shall first
present a relation for uniformly averaged matrix elements of a general linear operator, and
subsequently we shall derive corollaries yielding the fidelity of quantum
operations in a number of different cases.\\
\noindent \textit{Theorem}. For any linear operator $M$ on an
$n$-dimensional complex Hilbert space, the uniform average of
$|\langle \psi | M| \psi \rangle |^2$ over state vectors
$|\psi\rangle$ on the unit sphere $S^{2n-1}$ in $\mathbb{C}^n$ is
given by
\begin{equation}\label{theorem}
\begin{split}
& \int_{S^{2n-1}} |\langle \psi | M| \psi \rangle |^2 dV  \qquad \qquad \qquad \quad  (L)\\
& = \frac{1}{n(n+1)}\big[\text{Tr}(MM^\dagger)+|\text{Tr}(M)|^2\big]
\ \ (R)
\end{split}
\end{equation}
where $dV$ is the normalized measure on the sphere.\\
\noindent \textit{Proof}. A recent detailed proof of this theorem is given in \cite{dankert}, and it
is also readily verified by use of a recent result for the averages of general polynomials of state
vector amplitudes over the unit sphere \cite{t-spline}. Since both of these proofs draw on elaborate
mathematical results, we shall present an alternative more straightforward derivation.  We first show
that (\ref{theorem}) is valid if $M$ is Hermitian. A Hermitian matrix $M$ can be diagonalized such
that $UMU^{-1} = \Lambda$, where $\Lambda$ is a diagonal matrix with elements
$\lambda_1,\ldots,\lambda_n \in \mathbb{R}$ and $U\in U(n)$. The left-hand and right-hand sides of
\eqref{theorem}, denoted by $L$ and $R$ obey the conjugation invariance: $L(U M U^{-1}) = L(M),\ R(U M
U^{-1})=R(M)$ for any $U\in U(n)$. This follows for the
left-hand side, $L$, by a change of variables $\psi\mapsto U\psi$, and for $R$, it is simply a property of the trace. It thus suffices to show that $L=R$ for a real, diagonal matrix $\Lambda$.\\
\indent $L(\Lambda)$ is a homogeneous polynomial of degree 2 in the real variables
$\lambda_1,\ldots,\lambda_n$, and unitary invariance implies that it is invariant under the exchange
of any two $\lambda_i$ and $\lambda_j$, which is only possible if $L$ is on the form:
\begin{equation*}
L = a\text{Tr}(\Lambda^2)+b\left(\text{Tr}\Lambda\right)^2
\end{equation*}
where $a$ and $b$ are constants that may depend on $n$. We find that $a = b = \frac{1}{n(n+1)}$ by
picking two particularly simple matrices and observing that
\begin{equation*}
a\cdot n + b\cdot n^2 = L(I)=1
\end{equation*}
and
\begin{equation*}
a+b=L\left(
\begin{bmatrix} 1 & \, & \, \\ \, & 0  & \,\\ \, & \, & \ddots \end{bmatrix}\right)
= \int_{S^{2n-1}} |c_0|^4 dV = \frac{2}{n(n+1)},
\end{equation*}
where $|\psi\rangle = \sum c_j |j\rangle$ and the last equality follows by
evaluation of the integral. We have thus shown that $L(\Lambda)=R(\Lambda)$ and the theorem applies for any Hermitian matrix.\\
\indent We also have $L(A) = R(A)$ where $A$ denotes an anti-Hermitian matrix ($A^\dagger=-A$), cf.:
\begin{equation*}
L(A)=L(\tfrac{1}{i}iA) = L(iA) = R(iA) = R(\tfrac{1}{i}iA) = R(A)
\end{equation*}
The general matrix $M = \frac{M+M^\dagger}{2}+\frac{M-M^\dagger}{2} = S+A$ is decomposed as a sum of a
Hermitian and an anti-hermitian matrix, respectively, and the explicit results,
\begin{equation*}
\begin{split}
 L(S+A)&=  \int_{S^{2n-1}} |\langle \psi|S|\psi\rangle
+\langle \psi|A|\psi\rangle|^2 dV \\
& = \int_{S^{2n-1}} |\langle \psi|S|\psi\rangle|^2 + |\langle
\psi|A|\psi\rangle|^2+ \\ & \qquad \langle \psi|S|\psi\rangle\cdot
\underbrace{\left( \langle \psi |A^\dagger|\psi\rangle + \langle
\psi|A|\psi\rangle \right)}_0 dV\\
& = L(S)+L(A)
\end{split}
\end{equation*}
and
\begin{equation*}
\begin{split}
R(S+A) & = \frac{1}{n(n+1)}\lbrace \text{Tr}((S+A)(S-A)) + \\
& \qquad
(\text{Tr}S+\text{Tr}A)(\text{Tr}S-\text{Tr}A)\rbrace \\
&= R(S) + R(A)
\end{split}
\end{equation*}
prove the theorem $\blacksquare$

\noindent As noted above, the theorem has been known for a short time, but its first application has
been as an ingredient in noise estimation and quantum process tomography \cite{EAZ}. We shall here
apply the theorem to compute the average fidelity explicitly in various cases, where the actual
operation is known and fully characterized.

\noindent \emph{Average fidelity of a unitary transformation.} We first calculate the average fidelity
of a process where our, possibly controllable, physical interactions yield a unitary evolution $U$
whereas we intend to perform the unitary operation $U_0$ on the system. For transmission, storage and
retrieval, $U_0$ is the identity, and for quantum computing $U_0$ is the desired quantum gate. When
different initial states $|\psi\rangle$ are considered, the squared overlap between the actual outcome
$U|\psi\rangle$ and the desired final state $U_0|\psi\rangle$, averages to
\begin{equation}\label{fidunitary}
\begin{split}
F &=\int_{S^{2n-1}} |\langle \psi | M| \psi \rangle |^2 dV \\ &=
\frac{1}{n(n+1)}\big[\text{Tr}(MM^\dagger)+|\text{Tr}(M)|^2\big],
\end{split}
\end{equation}
with $M=U_0^\dagger U$. Since $M$ is unitary, $\text{Tr}(MM^\dagger)$ equals $n$, and if $M$
multiplies all states by a single phase factor, $|\text{Tr}(M)|^2$ equals $n^2$ and the fidelity is
unity. A general unitary matrix has eigenvalues of the form $\exp(i\phi_j)$, and $|\text{Tr}(M)|$ will
be smaller than $n$ if these phases differ. The "worst case" fidelity, i.e., the lowest fidelity
obtained for any state, is readily identified from the set of values $\phi_j$ \cite{WM}. The design of
composite pulses or smooth control theory pulses to implement gates with good robustness against
variations in atomic parameters is significantly facilitated with
our simple fidelity expression. \\
\indent Apart from the mean fidelity and the minimum, it might be
interesting to know the probability distribution for $F$. This is a
difficult problem, as witnessed by the probability for the integrand
in Eq.(1) for a Hermitian matrix $M$, studied originally by von
Neumann \cite{Neumann}.

\noindent \emph{Subspace averaged fidelity of a unitary
transformation.} In a number of quantum information scenarios,
auxiliary quantum levels are used to mediate the desired operations.
Qubits may for example be encoded in ground state atomic levels
which are coupled by Raman transitions via optically excited states
and two-qubit interactions may involve state-selective excitation to
states with large dipole moments and strong long-range interaction
\cite{Jaksch}. In these protocols, the auxiliary levels of the
quantum system are unpopulated before the process, and ideally, they
are also unpopulated after the process. But, if the fidelity is
evaluated as above by identifying $M=U_0^\dagger U$ on the complete
system Hilbert space, this may give far too pessimistic results.
Consider for example an atomic qubit which is initially and finally
with certainty in the ground hyperfine states, and no population has
leaked to the excited atomic states during the Raman process. If the
average in (1) is made over the full system Hilbert space, phases
acquired by amplitudes on the excited states cause a reduction in
the fidelity, even if the final state is the correct one for all
input qubit states. We should only average the fidelity over the
relevant input states. Since the final state, ideally, is also in
the qubit subspace, we thus consider the matrix
$M_{rel}=PU_0^\dagger UP$, where $P$ is the projection operator on
the relevant, quantum information carrying subspace. In matrix
notation, we thus address the square $n_{rel} \times n_{rel}$
submatrix of the full matrix $M$, and use our theorem for this
matrix.
\begin{equation}\label{fidrest}
\begin{split}
F & =\int_{S^{2n_{rel}-1}} |\langle \psi | M_{rel}| \psi \rangle |^2
dV \\ &=
\frac{1}{n_{rel}(n_{rel}+1)}\big[\text{Tr}(M_{rel}M_{rel}^\dagger)+|\text{Tr}(M_{rel})|^2\big],
\end{split}
\end{equation}
Note that if population leaks to the auxiliary levels, $M_{rel}$ is not unitary, and hence both terms
of the expression have nontrivial values. If a measurement assures that the final state does not
populate the complement to the relevant subspace, we get the average \emph{conditional} fidelity
$F_c=F/Q$, where $Q$, given by the same expression \eqref{fidrest} with $M_{rel}=PU^\dagger P U P$, is
the average acceptance probability of such a measurement.

\noindent \emph{Average fidelity for a general quantum operation.}
In a number of quantum information scenarios, ancillary quantum
systems are used to mediate the desired operations: quantum memory
protocols in a very explicit manner involve an extra quantum system,
quantum teleportation requires an extra entangled pair of systems,
and in quantum computing with trapped ions, a motional degree of
freedom is used to couple the particles. The ancillary systems are
ideally disentangled from the qubits before and after the process,
but in general they act as an environment and cause decoherence of
the quantum system of interest. This forces us to generalize the
formalism even further and take into account the general theory of
quantum operations, according to which density matrices are
transformed by completely positive maps. According to the Kraus
representation theorem, any completely positive trace-preserving map
$\mathcal{G}$ admits the representation
\begin{equation}
\mathcal{G}(\rho) = \sum_k G_k \rho G_k^\dagger
\end{equation}
where $\sum_k G_k^\dagger G_k = I_n$ is the $n \times n$ identity
matrix \cite{NC}. If there is only one $G_k$, this must be unitary
and we return to the case discussed above. The more general
expression allows interactions with an ancilla system followed by a
partial trace over that system, and it entails the effects of
dissipative coupling to the surroundings and of measurements on the
system, in which case the $G_k$ may represent the associated
projection operators and possible conditioned feed-back operations
applied to the system. It is a powerful feature of \eqref{theorem}
that it is easily extended to allow computation of the average
fidelity for protocols, which apply or which are subject to these
more general operations.

If the pure input state $\rho = |\psi\rangle \langle \psi|$ is mapped to the output state
$\mathcal{G}(\rho)$ the mean fidelity with which our operation yields a unitary transformation $U_0$
is
\begin{equation}\label{cor}
\begin{split}
F &= \int_{S^{2n-1}} \langle \psi | U_0^\dagger \mathcal{G}(|\psi\rangle\langle\psi|) U_0|\psi\rangle dV \\
&= \sum_k \int_{S^{2n-1}} |\langle \psi
|M_k|\psi\rangle|^2 dV\\
& = \frac{1}{n(n+1)} \Bigl\{ \text{Tr} \Bigl(\sum_k M_k^\dagger M_k \Bigr)+\sum_k |\text{Tr} (M_k)|^2
\Bigr\},
\end{split}
\end{equation}
where $M_k = U_0^\dagger G_k$, and $\{G_k \}$ are the Kraus operators for the map $\mathcal{G}$.

The Kraus representation is not unique as the map $\mathcal{G}$ might be represented by an alternative
set of operators $\{G_k'\}$, where $G_k' = \sum_j V_{kj} G_j$ and $V$ is a unitary transformation
among the Kraus operators \footnote{Note that if the set $\{G_k'\}$ contains more elements than
$\{G_k\}$, some zero operators are added to the set $\{G_k\}$ such that the two sets have the same
cardinality.}. We notice, however, that the first term in \eqref{cor} is the identity which is
unchanged, and since
\begin{equation*}
\begin{split}
\sum_k |\trace (M_k)|^2 &= \sum_k \trace(M_k^\dagger) \trace(M_k)
\\ &= \sum_{ij} \trace(M_j'^\dagger)\trace(M_i') \sum_k V_{jk}^\dagger
V_{ki} \\ &= \sum_j \trace(M_j'^\dagger) \trace(M_j') = \sum_k |\trace (M_k')|^2,
\end{split}
\end{equation*}
the fidelity is invariant under the transformation.

Eq. \eqref{cor} enables the calculation of the average fidelity of any quantum operation, as soon as
it has been put on the Kraus form. If a system for example undergoes dissipation, we can take this
into account together with other imperfections described in the previous sections, and from the
general solution of the associated Lindblad master equation we can determine the average fidelity of
our quantum state operation. By the procedure outlined in \cite{NC}, the general propagator for the
master equation can be systematically put on the Kraus operator form, and we also note the recent work
\cite{NHYMNM}, that explicitly  provides a Kraus form solution of the Lindblad master equation.

Let us demonstrate the formula \eqref{cor} on the depolarizing
channel, where a qubit state is unchanged with probability $p$ and
transformed into the random state $I_2/2$ with probability $1-p$.
This may occur as consequence of a continuous perturbation with a
rate $\gamma$ ($p=\exp(-\gamma t))$, or it may represent the outcome
of a teleportation process \cite{HHH}, \cite{Nielsen} where the
channel is not occupying an ideal maximally entangled state
two-qubit state $|\Psi\rangle$, but the mixed  state, $\rho_{Ent}
=p|\Psi\rangle\langle\Psi|+\frac{1-p}{4}I_4$. The transformation of
our qubit is represented by the action of four Kraus operators,
$M_0=\frac{\sqrt{3p+1}}{2}I_2,\
M_\chi=\frac{\sqrt{1-p}}{2}\sigma_\chi$, where $\sigma_\chi$ with
$\chi=x,y,z$ are the Pauli matrices. The mean fidelity for
preservation of the state under the depolarizing channel, or for the
teleportation of a qubit by the mixed state, follows readily from
\eqref{cor}, and
\begin{equation*}
F_{depolarized}=\frac{1}{6}(2+(3p+1))=\frac{1+p}{2},
\end{equation*}
as expected.

Another example is atomic decay with a decay rate $\Gamma$ between an upper and a lower qubit level,
where the Kraus map is
given by two operators, $M_0=\Bigl(\begin{smallmatrix}  1 & 0 \\
  0 & \exp(-\Gamma t/2) \\\end{smallmatrix}\Bigr),\
M_1=\Bigl(\begin{smallmatrix}
  0 & \sqrt{1-\exp(-\Gamma t)} \\
  0 & 0 \\
\end{smallmatrix}\Bigr)$.
In an analysis using Monte-Carlo Wave Functions \cite{DCM} these operators represent the no-jump and
the jump evolution, respectively, weighted by the corresponding time dependent probability factors.
The fidelity \eqref{cor} is readily evaluated with the result:
\begin{equation*}
F_{decay}=\frac{1}{6}(3+\exp(-\Gamma t)+2\exp(-\Gamma t/2)).
\end{equation*}

It is straightforward to apply our equations \eqref{fidunitary},
\eqref{fidrest} and \eqref{cor}, and this is a great advantage, in
particular for higher dimensional systems, where the explicit
integration over the space of initial states becomes cumbersome, and
where even the algebraic expressions in \cite{Nielsen}, \cite{BOSBJ}
become involved.

\noindent \textit{Average fidelity for general operations on
composite quantum systems}. A particular example of higher
dimensional systems is composite systems, as, e.g., a $K$-qubit or
$K$-qudit quantum register, where our formalism for not too large
$K$ can be used to determine the fidelity of more complex
operations, e.g., the achievements of error correcting codes
\cite{ecc}.

Let us consider here a simple example, where each qubit or qudit is subject to the same operation,
which may be either unitary or a Kraus form operation. This scenario is relevant for the teleportation
of a full quantum register by sequential or parallel teleportation of every individual bit, or simply
for the storage of a quantum register which suffers from independent decoherence mechanisms on every
bit. If the Kraus operators for a single qudit are denoted $M_k$,  the Kraus operators for $K$ qudits
are tensor product combinations of the $M_k$'s of the form $\mathcal{M}_{\vec{k}} =\bigotimes_i^K
M_{k_i}$, with $\vec{k}=(k_1,k_2,\ldots,k_K)$. It follows from \eqref{cor} that for the average over
the $n^K$ dimensional Hilbert space
\begin{equation*}
\begin{split}
F_{n^K} &= \frac{1}{n^K+1} + \frac{1}{n^K(n^K+1)} \sum_{\vec{k}}
|\text{Tr} (\mathcal{M}_{\vec{k}})|^2\\
& = \frac{1}{n^K+1} + \frac{1}{n^K(n^K+1)} \Bigl( \sum_k |\text{Tr} (M_k)|^2\Bigr)^K
\end{split}
\end{equation*}
where we used that  $\text{Tr}\left( \bigotimes_i^K M_{k_i}\right) =\prod_i^K \text{Tr} (M_{k_i})$.
Interestingly, using \eqref{cor}, we can express the sum $\sum_k |\text{Tr} (M_k)|^2$ in terms of the
single qudit fidelity $F_{n}$, and hence we obtain the relation between the $K$-qudit and the
single-qudit transformation fidelities:
\begin{equation}\label{qudits}
F_{n^K} = \frac{1}{n^K+1}\left( 1+((n+1)F_n-1)^K\right).
\end{equation}
We note that if $F_n=1$, also $F_{n^K}=1$, and due to \eqref{cor}, $F_n \geq 1/(n+1)$ for any map,
which consistently implies that $F_{n^K} \geq 1/(n^K+1)$. Contrary to what one might naively expect,
the fidelity for $K$ independent qudits is not just $F_n^K$, but it is given by the more complicated
expression \eqref{qudits}. This is due to the fact that the $K$'th power of the one qudit fidelity
formula only describes how fast qubit \textit{product states} decay on average, it does not average
over initial entangled states of the qudits in the register. Eq. \eqref{qudits} implies that for $F_n
= 1-\epsilon$, where $\epsilon \ll 1$, $F_{n^K}\sim 1- \frac{n^K}{n^K+1} \frac{n+1}{n} \cdot
K\epsilon$ which is smaller than $F_n^K \sim 1-K\epsilon$. For the special case of many qubits, $n=2$
and $K\gg 1$, we have the result $F_{n^K} \sim F_n^{3K/2}$. The result \eqref{qudits} and the limiting
values apply for any operation applied to the individual qudits, it only assumes that they are all
subject to the same individual dynamics. For few qudit systems one may well generalize the result to
the case of different operations on the individual components, if for example part of the qubits in a
quantum register are transferred to an interaction region by teleportation.

In summary, we have presented a derivation and shown various applications of  a simple and compact
expression for the average fidelity of general quantum operations. The problem has been addressed
previously, and alternative expressions, involving discrete sums over a large number of initial
states, have been discussed in the literature with emphasis on quantum process tomography, whereas, to
our knowledge, the expression involving only simple combinations of the trace of the relevant Kraus
operators has not been used in direct fidelity calculations. Our simple expression is a good starting
point for further analysis, e.g., of the achievements of error correcting codes \cite{ecc},
decoherence free subspaces \cite{dfs} and protection of quantum information by dynamical decoupling
\cite{Viola}. Our expression can also be handled and generalized analytically as illustrated by our
study of a $K$-qudit register, which provides insight into the scaling of errors which may have
applications in quantum error correction, the capacity of quantum channels, and the way that, e.g.,
communication with quantum repeaters \cite{repeaters} and entanglement distillation should optimally
be carried out. We apply an average over pure states, but we note that it is possible and also very
interesting to define fidelities for mixed initial states and to perform averages over mixed states
\cite{FPBTHC}. Since mixed states can be synthesized as the partial trace of pure states on a larger
system, we imagine that the accomplishments of operations on mixed states can also be addressed with
our formalism. The ability to restrict averages to subspaces may enable generalization of our
formalism to deal with non-uniform averages, assuming nontrivial prior probability distributions on
the Hilbert space. Also, in infinite-dimensional Hilbert spaces we may use our formulas, omitting the
$n(n+1)$ prefactor, to \emph{define} relative fidelity measures for any trace class operators. By
means of the regularized traces used for quantum field theory \cite{ntc}, it is even possible to
extend such measures canonically to certain non-trace class
operators.\\
\indent The authors gratefully acknowledge discussions with Uffe V.
Poulsen and Bent \O rsted. The project is supported by the European
Integrated Project SCALA and the ARO-DTO grant no. 47949 PHQC.


\begin{thebibliography}{99}

\bibitem{HHH} M. Horodecki, P. Horodecki and R. Horodecki, Phys. Rev. A \textbf{60}
1888 (1999)

\bibitem{Nielsen} M. A. Nielsen, Phys. Lett. A \textbf{303} 249
(2002)

\bibitem{BOSBJ} M. D. Bowdrey, D. K. L. Oi, A. J. Short, K. Banaszek, and J. A. Jones, Phys. Lett. A \textbf{294} 258
(2002)

\bibitem{BBM} E. Bagan, M. Baig and R. Munoz-Tapia, Phys. Rev. A \textbf{67} 014303
(2003)

\bibitem{NC} M. A. Nielsen and I. L. Chuang, Quantum Computation and
Quantum Information, Cambridge University Press, Cambridge (2000)

\bibitem{dankert} C. Dankert, \emph{Efficient Simulation of Random Quantum States and Operators}, Math. Thesis, University of Waterloo; quant-ph/0512217

\bibitem{t-spline} A. Ambainis and J. Emerson, \emph{Quantum t-designs: t-wise independence in the
quantum world}; quant-ph/0701126

\bibitem{EAZ} J. Emerson, R. Alicki and K. \.{Z}yczkowski, J. Opt. B: Quantum Semiclass. Opt. \textbf{7},
S347 (2006)

\bibitem{WM} J. Wesenberg and K. M\o lmer, Phys. Rev. A \textbf{68}
012320 (2003)

\bibitem{Neumann} J. von Neumann, Ann. of Math. Stat.
\textbf{12} 367 (1941)

\bibitem{Jaksch} D. Jaksch, J. I. Cirac, P. Zoller, S. L. Rolston, R. C\^{o}t\'{e}, and M. D. Lukin
Phys. Rev. Lett. {\bf 85}, 2208 (2000).


\bibitem{NHYMNM} H. Nakazato, Y. Hida, K. Yuasa, B. Militello, A. Napoli and A. Messina, Phys. Rev. A \textbf{74} 062113 (2006)

\bibitem{DCM} J. Dalibard, Y. Castin and K. M\o lmer, Phys. Rev.
Lett. \textbf{68}, 580 (1992)

\bibitem{ecc} A. Steane, Nature {\bf 399}, 124 (1999); E. Knill, Nature {\bf 434} 39 (2005).

\bibitem{dfs} P. Zanardi and M. Rasetti, Phys. Rev. Lett. \textbf{79}, 3306 (1997);
D. A. Lidar, I. L. Chuang, and K. B. Whaley, Phys. Rev. Lett.
\textbf{81}, 2594 (1998).

\bibitem{Viola} L. Viola, E. Knill, and S. Lloyd, Phys. Rev. Lett. {\bf 82}, 2417 (1999).

\bibitem{repeaters} L.-M. Duan, M. D. Lukin, J. I. Cirac and P. Zoller, Nature {\bf 414}, 413 (2001).

\bibitem{FPBTHC} E. M. Fortunato, M. A. Pravia, N. Boulant, G. Teklemariam, T. F. Havel, and D. G. Gory, J. Chem. Phys.
\textbf{116} 7599 (2002)

\bibitem{ntc} S. Paycha in Geometric methods for quantum field theory, Eds., H. Ocampo, S. Paycha and A. Reyes, World Scientific 2001



\end{thebibliography}
\end{document}